\documentclass[pdflatex,sn-mathphys-num]{sn-jnl}
\usepackage{amssymb}
\usepackage{amsthm}
\usepackage{amsmath}
\usepackage{mathtools}
\usepackage{bigints}
\usepackage{esint}
\usepackage{siunitx}
\usepackage{nicefrac}

\usepackage{hyperref}
\usepackage{lipsum}
\usepackage{nomencl}
\usepackage{comment}
\usepackage{import}

\usepackage{svg}
\usepackage{graphicx}
\graphicspath{{graphics/}}
\usepackage[section]{placeins}

\newcommand{\phase}{\phi}
\newcommand{\surf}{\Gamma}
\newcommand{\evolvsurf}{\Gamma(t)}

\newcommand{\vzero}{v_0}
\newcommand{\meanCurv}{H}

\newcommand{\normal}{\boldsymbol{n}}
\newcommand{\vnormal}{v_{\normal}}

\newcommand{\surfParam}{\boldsymbol{X}}

\newcommand{\kfac}{k_{fac}}
\newcommand{\klower}{k_{lower}}

\newcommand{\FCH}{\mathcal{F}_{CH}}
\newcommand{\FINT}{\mathcal{F}_{Int}}
\newcommand{\FHELF}{\mathcal{F}_{Helf}}
\newcommand{\eps}{\epsilon}
\newcommand{\FacCa}{Ca}
\newcommand{\invCa}{\frac{1}{\FacCa}}
\newcommand{\FacIn}{In}
\newcommand{\invIn}{\frac{1}{\FacIn}}
\newcommand{\kbend}{\kappa_{bending}}
\newcommand{\kbendh}{\kappa_{bending,h}}
\newcommand{\kbendPhi}{\kbend(\phase_1,..,\phase_N)}

\newcommand{\gradS}{\nabla_\surf}

\newcommand{\laplaceS}{\Delta_\surf}
\newcommand{\laplaceC}{\Delta_{\boldsymbol{C}}}

\newcommand{\intSt}[1]{\int_{\evolvsurf} #1\,d\evolvsurf}
\newcommand{\intSNull}[1]{\int_{\Gamma(0)} #1\,d\Gamma(0)}
\newcommand{\dt}{\partial_t}

\newcommand{\shapeOperator}{\mathcal{B}}

\newcommand{\rSurf}{r_{SH}}
\newcommand{\sphericalHarmonics}[2]{Y^{#1}_{#2}}
\newcommand{\argphi}{\varphi}
\newcommand{\argtheta}{\theta}

\newcommand{\qml}[2]{q^{#1}_{#2}}
 \newcommand{\ql}[1]{q_{#1}}
  \newcommand{\qlnorm}[1]{\tilde{q_{#1}}}

\title{Towards a two-scale model for morphogenesis - How cellular processes influence  tissue deformations}

\author[1]{\fnm{Lea} \sur{Happel}}\email{lea.happel@tu-dresden.de}
\author*[1,2,3]{\fnm{Axel} \sur{Voigt}}\email{axel.voigt@tu-dresden.de}
\affil[1]{\orgdiv{Institute of Scientific Computing}, \orgname{TU Dresden}, \orgaddress{\postcode{01062}, \city{Dresden} \state{Germany}}}
\affil[2]{\orgname{Center of Systems Biology Dresden (CSBD)}, \orgaddress{\street{Pfotenhauerstr. 108}, \postcode{01307}, \city{Dresden} \state{Germany}}}
\affil[3]{\orgdiv{Cluster of Excellence, Physics of Life (PoL)}, \orgname{TU Dresden}, \orgaddress{\street{Arnoldstr. 18}, \postcode{01307}, \city{Dresden} \state{Germany}}}

\abstract{We propose a two-scale model to resolve essential features of developmental tissue deformations. The model couples individual cellular behavior to the mechanics at tissue scale. This is realized by a multiphase-field model addressing the motility, deformability and interaction of cells on an evolving surface. The surface evolution is due to bending elasticity, with bending properties influenced by the topology of the cellular network, which forms the surface. We discuss and motivate model assumptions, propose a numerical scheme, which essentially scales with the number of cells, and explore computationally the effect of the two-scale coupling on the global shape evolution. The approach provides a step towards more quantitative modeling of morphogenetic processes.}

\begin{document}

\maketitle
\section{Introduction}

More than 100 years ago D'Arcy Thompson \cite{Tompson} pointed out correlations between biological forms and mechanical phenomena and postulated that biological shape diversity is revealing of the physical forces driving it. While this postulation today is confirmed, the details of the multiscale process of developmental tissue deformations, which couples individual cellular behavior to the mechanics at tissue scale are still largely unknown. Indeed, developmental tissue deformations require mechanical forces and during early multicellular development, these forces are often generated in effectively two-dimensional tissues \cite{Heisenberg_Cell_2013}. The large-scale motion during these processes emerges from collective tissue-mechanical properties of multicellular structures. For this reason various continuous theories have been proposed which consider coarse-grained material features and model shape evolution using physical conservation laws, e.g., hydrodynamic theories of active thin shells \cite{salbreux2012actin,da2022viscous,de2023cell,khoromskaia2023active}, active polar or nematic surfaces \cite{julicher2018hydrodynamic,morris2019active,salbreux2022theory,al2023morphodynamics,Nitschke_PRSA_2025} or models for active fluid deformable surfaces \cite{torres2019modelling,reuther2020numerical,krause2023numerical,Porrmann_PF_2024}. While huge progress has been made with respect to model derivation and numerical simulations \cite{salbreux2022theory,da2022viscous,nitschke2025beris,torres2020approximation,nestler2019finite,Porrmann_PF_2024}, so far, these models are still unable to quantitatively predict the dynamics of morphogenetic processes. A possible step towards more quantitative modeling is to take processes on the cellular scale stronger into account. Various "discrete" models exist which focus on the cellular scale, e.g., \cite{kim2021embryonic,fuhrmann2024active,runser2024simucell3d,Ouzeri2025.03.23.644792}.  We propose a two-scale coupling mechanism, which intervenes the cellular and the tissue scale. Instead of a continuous theory with coarse-grained material properties we resolve the individual cells in the effectively two-dimensional tissue, and consider their deformability, their motility and their interaction with neighboring cells. These cellular processes form the tissue and determine the mechanical forces required for tissue deformation. However, the coupling is in both directions, the shape of the two-dimensional tissue also influences the cells. In order to account for these couplings we adapt mechanical phenomena known for thin crystalline sheets. Out of plane deformations at topological defects \cite{PhysRevA.38.1005,lehtinen2013atomic,Benoit_Marechal_MM_2024} is one example. Also in tissues topological defects can be defined. Various efforts relate to orientational defects \cite{saw2017topological,Nitschke_PRSA_2020,guillamat2022integer,hoffmann2022theory,MaroudasSacks_NP_2021,maroudas2025mechanical}. However, also positional defects play a role. Presumably regions with cells with five or seven neighbors have different mechanical properties than those with cells with only six neighbors. Higher-order vertices or multicellular rosettes also have been shown
to have an effect on the mechanics of the tissue \cite{Bietal_PRX_2016}. Such defects greatly influence the rigidity of the epithelial tissue. Neighbor exchanges between the cells determine the flow of the tissue and are essential in many important aspects of morphogenesis. The two-scale coupling thus essentially allows to study the influence of cellular processes on the tissue scale. This in principle would also allow to consider genetic and mechano-chemical processes, however, we here refrain from such attempts and only model the cells as deformable active Browning particles, neglecting any sub-cellular properties. We consider a multiphase-field model to resolve each individual cell, which has been shown to provide a reliable modeling approach, at least in flat space \cite{Camley_PNAS_2014,Mueller_PRL_2019,Loewe_PRL_2020,Wenzel_PRE_2021}, next to, e.g., vertex models \cite{Farhadifaretal_CB_2007,Bietal_PRX_2016,rozman2025vertex}, which operate on the same spatial scales. However we consider this model on an evolving surface, with the evolution defined by the two-dimensional tissue. We therefore extend previous approaches of multiphase-field models on stationary surfaces \cite{Happel_EPL_2022,Happel_PRL_2024} and combine them with Canham/Helfrich models to consider the bending properties of the tissue.  

The outline of the paper is as follows: In Section \ref{sec:2} we introduce the mathematical model, motivate and justify all assumptions, introduce notation and the system of equations and explain how they are discretised. In Section \ref{sec:3} we discuss results, starting from known phenomena of classical continuous models and comparisons with previous studies for cells on stationary geometries, before the full two-scale coupling is fully explored. Finally we discuss the results and draw conclusions in Section \ref{sec:4}.      

\section{Mathematical model}
\label{sec:2}

\subsection{General description and model assumptions}
We make the following assumptions: The thickness of the monolayer of epithelial tissue is assumed to be constant and small compared to its lateral extension. This allows to neglect the thickness and to model the tissue as a two-dimensional surface. We assume this surface to be closed, have constant area, constant enclosed volume and have properties of an elastic film, which can be solid or fluid like, depending on the properties of the tissue. We assume the film to only resist bending and consider a Canham/Helfrich energy. Minimizing this energy under the mentioned constraints using a $L^2$-gradient flow essentially determines the shape change and with it the movement of the surface in normal direction. So far we have a fully continuous description of the epithelial tissue. This description is well known from models of fluid membranes. In this context the microscopic constituents, the lipids, are only considered in a coarse-grained manner. Their properties in tangential direction are either considered implicitly \cite{Canh70,Helfrich_ZNC_1973} or more recently if membrane viscosity is taken into account also explicitly \cite{arroyo2009relaxation,reuther2015interplay,reuther2018erratum,torres2019modelling}. However, also these models are continuous and operate on the same length scale as the Canham/Helfrich energy. In the context of epithelial tissue the surface consists of interacting cells. A coarse-grained description of such a material is a challenging task even in flat space \cite{grossman2022instabilities,claussen2025mean,josep2024hydrodynamics,nejad2025coarse} and most studies therefore resolve each individual cell. Thereby each cell is modeled as an active deformable object. We consider each cell to have constant area, therefore neglect cell growth or shrinkage, and consider a line tension to energetically evolve its shape. Activity is introduced via a convective term. This, relatively simple description of a cell, can be seen as a generalization of an active Brownian particle \cite{Loewe_PRL_2020}, and is successfully used to model collective motion in flat space \cite{Wenzel_PRE_2021,monfared2025multi} and on curved surfaces \cite{Happel_EPL_2022,Happel_PRL_2024}. In our context the cell builds part of the surface and a confluent collection of cells, with cell-cell interactions, forms the surface. The constant area of cells thus leads to the constraint of a constant area of the surface. We further assume all cells to have equal size and properties. Within this setting we have a two-scale model which is continuous with respect to bending and thus movement in normal direction, but "discrete", resolving each individual cell, in tangential direction. From experimental and theoretical studies of monolayers of epithelial tissue in flat space it is well known that topological features resulting from the cell arrangements have strong implications on the mechanical properties of the tissue \cite{Bietal_PRX_2016,Jain_SR_2023}. We here consider a computationally accessible topological feature, the number of neighbors of a cell, to account for this dependency. In order to couple the "discrete" cellular scale to the continuous scale of the surface we make the bending rigidity to depend on the number of neighbors. As for two-dimensional crystalline materials, which buckle at defects \cite{PhysRevA.38.1005,lehtinen2013atomic,Benoit_Marechal_MM_2024}, the bending rigidity is reduced for cells for which the number of neighbors deviate from six, where six corresponds to the optimal hexagonal packing in flat space. This provides additional coupling between the different scales of the two-scale problem. The coupling is in both directions, the cellular arrangement influences the bending rigidity and thus the shape change of the continuous surface and the surface evolution influences the shape and movement of each cell, as the cells are confined to the surface. These couplings allow to explore how activity, which is induced on the "discrete" cellular scale, lead to morphological changes of the surface on the continuous scale. 

\subsection{Notation}

The considered two-scale model requires basic notation from differential geometry and geometric partial differential equations. We follow the same notation as in \cite{Bachini_JFM_2023}. The basic parts are repeated for convenience. We consider a time dependent smooth and oriented surface $\surf = \surf(t)$ without boundary, embedded in $I\!\!R^3$ and given by a parametrization $\surfParam$. The enclosed volume is denoted by $\Omega = \Omega(t)$. We denote by $\normal$ the outward pointing surface normal, the shape operator is $\shapeOperator = -\nabla_{\mathbf{P}} \normal$ with $\nabla_{\mathbf{P}}$ the tangential derivative, and the mean curvature is $\meanCurv= \text{tr}\shapeOperator$. With this notation the mean curvature of the unit sphere is $\meanCurv = -2$. We further consider the covariant derivative $\nabla_\Gamma$ and the covariant divergence $\nabla_\Gamma \cdot$, with $\Delta_{\Gamma} = \nabla_\Gamma \cdot \nabla_\Gamma$ the Laplace-Beltrami operator. $\Delta_C$ denotes a componentwise application of $\Delta_\Gamma$. The model in the following should be considered in dimensionless form.

\subsection{Model formulation}

The simplest mechanical model for a two-dimensional surface that only resists bending is the Canham/Helfrich model which is based on the energy
\begin{align}
\FHELF &= \intSt{\frac12 \kbend \meanCurv^2}
\end{align}
where $\kbend$ is the bending rigidity and the spontaneous curvature is set to zero \cite{Canh70,Helfrich_ZNC_1973}. With constraints on constant area and constant enclosed volume the corresponding evolution equations read
\begin{align}
    \vnormal &= - \frac{\delta \FHELF}{\delta \surfParam} \cdot \normal - \lambda_{vol} + \lambda_{area} \meanCurv \label{eq:helf1} \\
   -\lambda_{area} \intSt{\vnormal\meanCurv} &= \intSNull{1} -\intSt{1}\label{eq:helf2} \\
    \lambda_{vol} \intSt{\vnormal}  &= \frac{1}{3}\intSNull{\langle \surfParam, \normal \rangle} - \frac{1}{3}\intSt{\langle \surfParam, \normal \rangle}\ \label{eq:helf3}
\end{align}
where $\vnormal$ is the magnitude of the velocity of the surface in normal direction, $\lambda_{vol}$ and $\lambda_{area}$ are Lagrange multipliers (ensuring conservation of the initial volume and area, similar to the technique used in \cite{Krause_JoTRSI_2024}), and
\begin{align}
     \frac{\delta \FHELF}{\delta \surfParam} &= \laplaceS(\kbend \meanCurv) \normal -\kbend \meanCurv \Big(||\shapeOperator||^2-\frac{\meanCurv^2}{2}\Big) \normal \label{eq:varFHelf}. 
\end{align}

\begin{figure}[htb]
    \includegraphics[width=\linewidth]{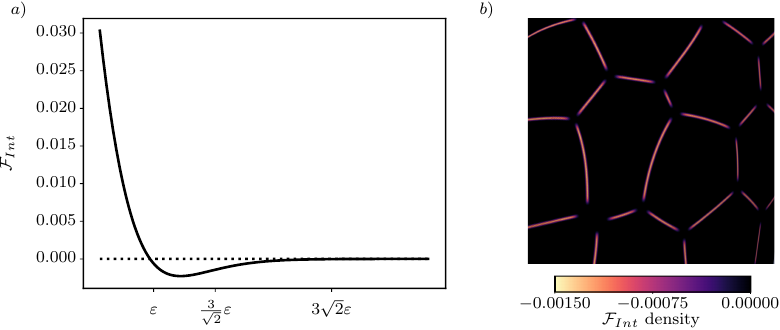}
    \caption{Short range interaction potential as a function of the distance between the $0$-levelsets of two cells a) and in a computational realization highlighting the short range interaction b). The used parameters are $\FacIn=0.05$, $a_{rep}=0.0625$ and $a_{att}=0.48$. For a phase-field with the stable phases $-1.0$ and $1.0$ the width of the interface is approximately $3\sqrt{2\eps}$ (obtained from the equilibrium tanh-profile). This point is marked explicitly in a). The underlying cell contours for b) can be seen in Figure \ref{fig:kbend} b).}
    \label{fig:interaction}
\end{figure}

This classical problem is supplemented by a surface multiphase-field model to account for cell arrangements which form the surface. We consider phase-field variables $\phase_i(\mathbf{x},t)$, one for each cell, with $\mathbf{x}$ defined on the surface $\surf$. Values of ${\phase_i=1}$ and ${\phase_i=-1}$ denote the interior and exterior of a cell, respectively. The cell boundary is implicitly defined as the zero-level set of $\phase_i$. We consider a surface Cahn-Hilliard-like energy
\begin{align}
     \FCH = \invCa \sum\limits_{i=1}^N \intSt{ \frac{1}{G(\phase_i)} \left(\frac{\eps}{2}\|\gradS\phase_i\|^2 + \frac{1}{\eps}W(\phase_i)\right)},
\end{align}
where $N$ denotes the number of cells, $Ca$ is the capillary number, measuring the deformability of the cells, $G(\phase_i) = \tfrac{3}{2}|1 - \phase_i^2|$ is a de Gennes factor which helps to keep ${-1 \leq \phi_i \leq 1}$ \cite{Salvalaglio_MMAS_2021}, $W(\phase_i) = \tfrac14 (1 - \phase_i^2)^2$ is a double well potential and $\eps$ a small parameter determining the width of the diffuse interface. Cell-cell interactions are considered by an interaction energy
\begin{align}
\FINT =\invIn \sum\limits_{i=1}^N \sum\limits_{j\neq i}\intSt{{a}_{rep} (\phase_i +1)^2(\phase_j+1)^2 - {a}_{att} W(\phase_i) W(\phase_j)} \label{eq_FINT}
\end{align}
where $In$ is the strength of interaction and $a_{rep}$ and $a_{att}$ model repulsive and attractive components, respectively. The first term in eq. \eqref{eq_FINT} penalizes overlap of the interior of the cells and the second favors overlap of the diffuse interfaces. The last can best be seen considering the equilibrium condition $\frac{\eps}{2} \|\gradS \phase_i\|^2 \approx \frac{1}{\eps} W(\phase_i)$ resulting from the $\tanh$-profile of $\phase_i$, see \cite{Guetal_JCP_2014}. The form is well established in flat space and also has already been considered on curved surfaces \cite{Happel_PRL_2024}. Adjusting the parameters $a_{rep}$ and $a_{att}$ allows to construct a short-range interaction potential within the diffuse interface, see Figure \ref{fig:interaction}, demonstrating the consistency with other approaches directly implementing an interaction potential \cite{Marth_IF_2016,Wenzel_JCP_2019}.

The evolution equation for each $\phase_i$ reads
\begin{align}
    \dot{\phase}_i - \phase_i \vnormal \meanCurv + \vzero (\mathbf{e}_i \cdot \gradS \phase_i) &= \laplaceS \frac{\delta (\FCH + \FINT)}{\delta \phase_i}
     \label{eq:EvoPhi}
\end{align}
which considers conservation of mass and the transport formula 
$$\frac{d}{dt} \int_{\Gamma(t)} \phase_i d \Gamma(t) = \int_{\Gamma(t)} \dot{\phase}_i  d \Gamma(t) + \int_{\Gamma(t)} \phase_i \nabla_\Gamma \cdot (\vnormal \normal) d \Gamma(t),$$ 
see \cite{dziuk2013finite}. Thereby, $\dot{\phase}_i = \partial_t \phase_i + \nabla_{\mathbf{w}} \phase_i$ is the material derivative with ${\nabla_\mathbf{w} \phase_i = (\nabla_\Gamma \phase_i, \mathbf{w})}$ and the relative material velocity $\mathbf{w} = \vnormal \normal - \partial_t \surfParam$, see \cite{Nitschke_JGP_2022,Bachini_JFM_2023}, and ${- \phase_i \vnormal \meanCurv = \phase_i \nabla_\Gamma \cdot (\vnormal \normal)}$. We follow an Lagrangian-Eulerian perspective, Lagrangian in normal direction providing the coupling with the normal velocity $\vnormal$ and Eulerian in tangential direction. The other terms are as in \cite{Happel_EPL_2022} and contain the self-propulsion strength $\vzero$ and direction $\mathbf{e}_i = (\cos{\theta_i}, \sin{\theta_i})$ considered in a local coordinate system for the tangent plane at the center of mass of cell $i$. For consistency we define one of the orthonormal vectors of this coordinate system with respect to the projected direction $\mathbf{e}_i$ of the previous time step. The angle $\theta_i$ is controlled by rotational noise ${d \theta_i(t) = \sqrt{2D_r} d W_i(t)}$, with diffusivity $D_r$ and a Wiener process $W_i$. The variational derivatives read:
\begin{align}
    \frac{\delta \FINT}{\delta \phase_i} &= \invIn \sum\limits_{j\neq i} \left( 4 a_{rep}(\phase_i+1)(\phase_j+1)^2-\frac12 a_{att}(\phase_i(\phase_i^2-1))(\phase_j^2-1)^2\right) \label{eq:varDivFINT} \\
        \frac{\delta \FCH}{\delta \phase_i}&= \invCa \left(\frac{1}{G(\phase_i)}\frac{1}{\eps}W'(\phase_i) - \frac{\eps}{G(\phase_i)}\laplaceS \phase_i+\left(\frac{1}{G(\phase_i)}\right)'\left(\frac{1}{\eps}W(\phase_i) - \frac{\eps}{2}\|\gradS\phase_i\|^2 \right)\right) ,
         \label{eq:varDivFCH}
\end{align}
where the last can be simplified using the equilibrium condition $\frac{\eps}{2} \|\gradS \phase_i\|^2 \approx \frac{1}{\eps} W(\phase_i)$ to obtain the approximation \cite{Salvalaglio_MMAS_2021} 
\begin{align}
        \frac{\delta \FCH}{\delta \phase_i}&= \invCa \left(\frac{1}{G(\phase_i)}\frac{1}{\eps}W'(\phase_i) - \frac{\eps}{G(\phase_i)}\laplaceS \phase_i\right). \label{eq:varch}
\end{align} 

For $\vzero = 0$ and only one cell $\phase$ the resulting system of equations is a surface Cahn-Hilliard equation on a bendable surface. It relates to phase-field approximations of a Jülicher-Lipowsky model for two-component vesicles \cite{caetano2023regularization,Bachini_JFM_2023,Sischka_Preprint_2025}. However, there are conceptional differences. These models only consider one spatial scale and a bending rigidity $\kbend = \kbend(\phase)$ with a functional dependency on $\phase$. This not only requires to consider $\FHELF + \FCH$ instead of $\FHELF$ in eqs. \eqref{eq:helf1} - \eqref{eq:helf3}, but also  
$\FHELF + \FCH$ instead of $\FCH$ in eq. \eqref{eq:EvoPhi}. In our context we can make use of the two-scale character. Due to separation of scales we question the necessity of $\FCH$ and $\FINT$ to be considered in eqs. \eqref{eq:helf1} - \eqref{eq:helf3}. Both are essentially only nonzero within the vicinity of the cell interfaces and thus on a small scale, which is not relevant for the large scale surface evolution. We therefore neglect these terms in eqs. \eqref{eq:helf1} - \eqref{eq:helf3}. Instead of a functional dependency of the bending rigidity $\kbend = \kbendPhi$ on all phase-field variables, which would require to consider $\delta \FHELF / \delta \phase_i$ in eq. \eqref{eq:EvoPhi}, we interpret $\kbend$ as a coarse-grained parameter addressing only topological features of the monolayer. The implicit dependency of these features on $\phase_i$ will be neglected in eq. \eqref{eq:EvoPhi}, the same holds for the resulting spatial dependency in eqs. \eqref{eq:helf1}-\eqref{eq:helf3}. Considering these approximations the coupled system eqs. \eqref{eq:helf1}-\eqref{eq:varFHelf}, \eqref{eq:EvoPhi}, \eqref{eq:varDivFINT} and \eqref{eq:varch} remain without further modification. 
\begin{figure}[!hbt]
\includegraphics[width=\linewidth]{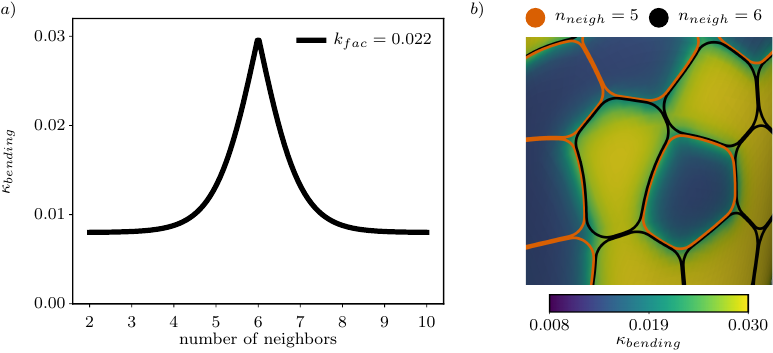}
    \caption{Bending rigidity $\kbend$ as a function of the number of neighbors $n_{neigh}$ for  $\kfac=0.022$ and $\klower=0.008$, shown as graph in a) and a numerical realization showing the cell contours color-coded by the number of neighbors and the values of $\kbend$ in b).}
    \label{fig:kbend}
\end{figure}

Various examples exist in which material properties are influenced by topological defects resulting from the underlying microstructure. Such features are well studies in flat space \cite{Jain_SR_2023,Jain_PRR_2024}, but have also been found in spherical surfaces \cite{Happel_EPL_2022}. Related to these features is the number of neighbors of a cell. Due to Euler's polyhedron formula, for the topology of a sphere there must be cells with the number of neighbors deviating from six. The neighbor distribution and the optimal arrangement of interacting cells on a sphere are a well studied problem. Depending on the number of cells, disclinations, isolated cells with five or seven neighbors, dislocations, pairs of five- and seven-fold disclinations, or grain boundary scares, chains of five- and seven-fold disclinations, emerge. Most studies consider fixed, e.g., circular cells, but the same phenomena can also be found for deformable cells \cite{Happel_EPL_2022}. Material properties differ at these topological defects, they are sources for weakness. We therefore modify the bending rigidity for these cells. The considered dependency reads 
\begin{align}
   \kbend(\mathbf{x},t) = \kfac \left( \tanh{ \left( -|n_{neigh}(\mathbf{x},t) - 6|  \right)} + 1 \right) + \klower \label{eq:kbend}
\end{align}
with parameters $\kfac$ and $\klower$ determining the maximal values for cells with six neighbors and the minimal values which is approached for increasing deviations from six. $n_{neigh}(\mathbf{x},t)$ is the number of neighbors of the cell, which center of mass is closest to position $\mathbf{x}$ at time $t$. We call two cells neighbors if their diffused interfaces overlap, so if they are able to interact. To be precise we follow \cite{Monfared_eLife_2023} and say two cells are neighbors if ${\{\phase_i|\phase_i>-0.5\}\cap \{\phase_j|\phase_j>-0.5\} \neq \emptyset}$. The dependency of $\kbend$ on $n_{neigh}$ is phenomenological and chosen similar to the discrete defect localization approach in \cite{Aland_MMS_2012}, where it was shown to lead to qualitatively similar results than variational based approaches \cite{Aland_MMS_2012,Benoit_Marechal_MM_2024}. The functional form and a realization for selected cells in a cell monolayer after numerical smoothing, see Section \ref{sec:2.4}, are shown in Figure \ref{fig:kbend}. 

\subsection{Discretization} \label{sec:2.4}

The overall system of geometric and surface partial differential equations is solved by surface finite elements in space \cite{dziuk2013finite,nestler2019finite} and classical finite differencing in time. We consider an operator splitting approach to decouple the Lagrangian movement in normal direction and the Eulerian evolution in tangential direction in each time step. For each subproblem we build on established approaches.

In order to numerically solve eqs. \eqref{eq:helf1}-\eqref{eq:helf3} we follow \cite{Barrett_JCP_2008}. As in \cite{Bachini_JFM_2023, krause2023numerical} we supplement these equations with an additional equation for the time evolution of the parametrization. We consider the initial surface given $\surfParam(0) = \surfParam_0$ and solve
\begin{align}
    \dt \surfParam \cdot \normal &= \vnormal \label{eq:reg1} \\
    \meanCurv \normal &= \laplaceC \surfParam \label{eq:reg2} .
\end{align}
This generates a tangential mesh movement that maintains the shape regularity and additionally provides an implicit representation of the mean curvature $\meanCurv$ \cite{Barrett_JCP_2008}. Eqs. \eqref{eq:helf1}-\eqref{eq:helf3} and eqs. \eqref{eq:reg1} and \eqref{eq:reg2} are solved together. We approximate the smooth surface $\Gamma(t)$ by a discrete $k$-th order approximation $\Gamma_h^k(t)$ \cite{Praetorius_ANS_2022} and consider each geometric quantity, like the normal $\normal_h$, the shape operator $\shapeOperator_h$ and the inner product $(\cdot,\cdot)_h$ with respect to $\Gamma_h^k(t)$. In the following we drop the index $k$. We define the discrete function spaces by $V_l(\Gamma_h) = \{\psi \in C^0(\Gamma_h)|\psi_{|T} \in P_l(T) \}$ with $P_l$ the space of polynomials and $T$ the element of the surface triangulation. We define $\mathbf{V}_l(\Gamma_h) = [V_l(\Gamma_h)]^3$ as space of discrete vector fields and consider $\surfParam_h \in \mathbf{V}_2(\Gamma_h)$ and $H_h, \kbendh \in V_2(\Gamma_h)$. For a detailed weak and finite element formulation we refer to \cite{Bachini_JFM_2023,krause2023numerical}. In order to obtain the same numerical properties and convergence behavior as in these approaches we smooth $\kbendh$ by solving one pseudo time step of $\partial_t \kbendh = \laplaceS \kbendh$.  

For solving eqs. \eqref{eq:EvoPhi}, \eqref{eq:varDivFINT} and \eqref{eq:varch} we extend the approach introduced for the problem in flat space \cite{Praetorius_NIC_2017} and considered on fixed curved surfaces \cite{Happel_EPL_2022,Happel_PRL_2024} to evolving surfaces. Based on $\Gamma_h(t)$, which serves as a macro mesh we consider $N$ different refinements $\Gamma_{h,i}(t)$ to resolve the $N$ phase-field variables $\phase_i$. This multimesh approach \cite{Praetorius_NIC_2017} allows to solve the $N$ problems in parallel. Only interactions between cells require communication, which is done on the only virtually available common finest mesh of the different refinements. In flat space this approach is shown to scale with the number of cells $N$ \cite{Praetorius_NIC_2017}. In our context the approach also resamples the two-scale nature of the problem. All geometric properties of the surface are only communicated on the common macro mesh $\Gamma_h(t)$. We split the fourth order systems into coupled systems of second order problems by defining $\mu_i = \delta (\FCH + \FINT) / \delta \phase_i$ and consider $\phase_{i,h}, \mu_{i,h} \in V_2(\Gamma_{h,i})$. A detailed weak and finite element formulation on stationary surfaces can be found in \cite{Happel_PRL_2024}. For the extension to evolving surfaces we follow \cite{Bachini_JFM_2023}. We essentially consider the relative material velocity as $\mathbf{w}_h^{n}$ in the material time derivative. 

With these two subproblems set the overall algorithm starts with an initial parametrization $\surfParam_h(0)$, phase-field variables $\phase_{i,h}(0)$ and chemical potentials $\mu_{i,h}(0)$ and evolves in time from $t^n$ to $t^{n+1}$ by first computing $\kbendh^n$ using $\phase_{i,h}^n$ and solving the geometric evolution problem for ${\vnormal}^{n+1}$, $\lambda_{area}^{n+1}$, $\lambda_{vol}^{n+1}$, $\surfParam_h^{n+1}$ and $\meanCurv_h^{n+1}$ and afterwards solve the system of surface partial differential equations for $\phase_{i,h}^{n+1}$ and $\mu_{i,h}^{n+1}$ using ${\vnormal}^{n+1}$, $\surfParam_h^{n+1}$ and $\meanCurv_h^{n+1}$. The approach is implemented in AMDiS \cite{AMDiS:2.10,Vey_CVS_2007,Witkowski_ACM_2015} which is based on the DUNE library \cite{Bastian_CMA_2021, Sander_Dunebook_2020}. For the surface approximation DUNE-CurvedGrid \cite{Praetorius_ANS_2022} together with DUNE-AluGrid \cite{Alkämper_ANS_2016} is used. For $v_0 = 0$, only one phase $\phase$ and a stronger constraint on the conservation of area the approach is shown to converge for $e_{\surfParam} = \|\surfParam_h - \surfParam \|_{L^\infty(L^2(\Gamma_h))}$ and
$e_\phase = \|\phase_h(\surfParam_h) - \phase(\surfParam)\|_{L^\infty(L^2(\Gamma_h))}$ experimentally with optimal order \cite{Bachini_JFM_2023,Sischka_Preprint_2025}. We relay on these results and refrain from further convergence studies, which, due to the complexity of the problem for $N \gg 1$, also become unfeasible.  

Further technical details and the considered parameters are listed in the Appendix.
\FloatBarrier
\section{Results} \label{sec:3}
\subsection{Comparison with classical Canham/Helfrich model}
\begin{figure}[!hbt]
    \centering
    \includegraphics[width=0.6\linewidth]{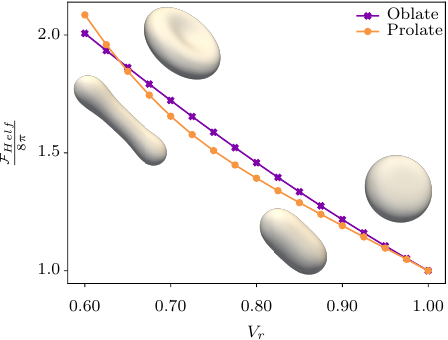}
    \caption{Rescaled bending energy $\nicefrac{\FHELF}{(8\pi)}$ of minimal energy configurations as function of reduced volume $V_r$ together with sample shapes for $\kbend =1.0$. The sample shapes correspond to ${V_r=0.65}$ (left) and $V_r=0.9$ (right) and show the oblate (top) and prolate (bottom) configuration.}
    \label{fig_Seifert}
\end{figure}

We first test the problem for consistency and set $v_0 = 0$. For $\kbend = const$ the shape evolution reduces to the classical Canham/Helfrich problem \cite{Canh70,Helfrich_ZNC_1973}. The preferred shapes arise from a competition between the Canham/Helfrich energy and the geometrical constrains on fixed surface area and fixed enclosed volume. These shapes of lowest energy can be classified in phase diagrams, which in the considered parameter regime show two stable branches of prolate and oblate shapes \cite{Seifert_1997}. Figure \ref{fig_Seifert} shows the computed diagram together with sample shapes for selected values of the reduced volume, which denotes the scaled quotient of the enclosed volume and the surface area $V_r = 6 \sqrt{\pi} |\Omega| / |\surf|^{3/2}$. A reduced volume $V_r = 1$ thus corresponds to a sphere. The phase diagram agrees with \cite{Seifert_1997}, where the shapes are computed in an axisymmetric setting. 

\subsection{Consistency tests for passive system}
\begin{figure}[b]
    \includegraphics[width=\linewidth]{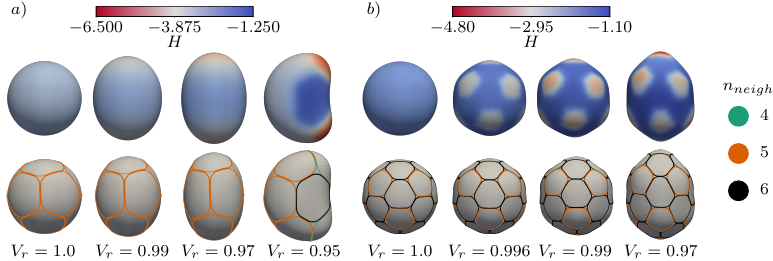}
    \caption{Equilibrium energy configurations for selected reduced volumes $V_r$ and $N=12$ a) and $N=32$ b). Top row shows the mean curvature $\meanCurv$ and bottom row the contours of the cells, color coded according to the number of neighbors.}
    \label{fig_Virus}
\end{figure}
\begin{figure}[!tbh]
    \includegraphics[width=\linewidth]{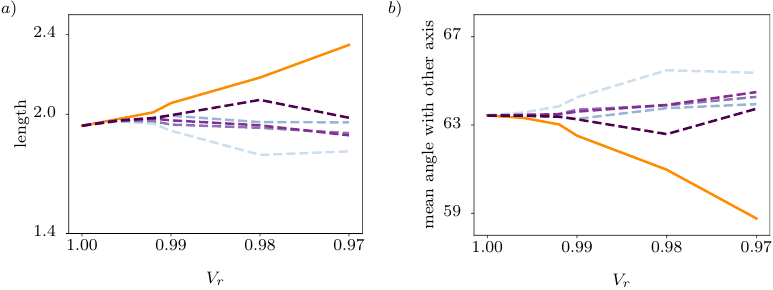}
    \caption{Symmetry breaking for $N= 32$. Reducing the reduced volume $V_r$ leads to truncated icosahedral shapes for $V_r \gtrsim 0.99$ (all six axis connecting opposed cells with 5 neighbors behave similar) and prolate like configurations for $V_r \lesssim 0.99$ (one axis (solid, orange) deviates from the others (dashed). Shown is the distance between the cells a) and the mean angle with the other axis b) as functions of the reduced volume $V_r$.}
    \label{fig_quant_axis}
\end{figure}

In the next step we are interested in how these shapes deform if the surface consists of cells and the proposed form $\kbend$ in eq. \eqref{eq:kbend} is used. We consider two cases $N = 12$ and $N = 32$. The optimal arrangements of 12 and 32 interacting cells on a sphere are know. Finding these arrangements is related to the classical Thomson problem for the interaction of points on a sphere \cite{thomson1904xxiv} and the Tammes problem, which asks for the positions such that the minimum distance between points is maximized, which is equivalent to the optimal packing of circles on the sphere \cite{1987AcCrA..43..612T}. For $N=12$ the results are known and identical. The points reside at the midpoints of the faces of a regular dodecahedron. For $N=32$ the minimal energy configurations form a truncated icosahedron. Also in this situation the topology of the minimal configurations, which resamples a classical soccer ball, agrees for both problems. The robustness of these configurations with respect to the specific interaction potential has also been demonstrated numerically, see \cite{Backofen_PRE_2010,Backofen_MMS_2011} for an approach based on a surface phase-field crystal model. We therefore expect to find these configurations also for our problem. As for an initial sphere the shape remains fixed, due to the area and volume constraint, the problem significantly simplifies and has already been solved in \cite{Happel_EPL_2022}. Indeed for these special cases the expected configurations are obtained. Starting from these configurations we now reduce the enclosed volume, keep the surface area fixed and relax the system by solving the full problem with $v_0 = 0$. While for $N=12$ all cells have initially 5 neighbors and thus $\kbend = const$, for $N=32$ we have 12 cells with five neighbors and 20 cells with six neighbors and thus variations in $\kbend$. We therefore expect to see changes in the emerging equilibrium configurations. Figure \ref{fig_Virus} shows the obtained minimal energy configurations for selected values of the reduced volume $V_r$. For $N=12$, see Figure \ref{fig_Virus} a), above some threshold for $V_r \approx 0.96$ the obtained shapes resample the equilibrium shapes of the prolate branch from Figure \ref{fig_Seifert}. This is a consequence of the definition of $\kbend$ in eq. \eqref{eq:kbend} which is homogeneous if the number of neighbors is constant for all cells. However, below this threshold the influence of the shape on the cells becomes apparent. Due to the two emerging strong curvature regions of the prolate the cells deform and rearrange, which leads to variations in the number of neighbors and thus heterogeneity in $\kbend$. This breaks the symmetry and forms a new (local) equilibrium shape not seen in Figure \ref{fig_Seifert}. For $N=32$, see Figure \ref{fig_Virus} b), the inhomogeneity in $\kbend$ leads to symmetry breaking for all $V_r < 1.0$. Above some threshold, $V_r \approx 0.99$, the surface approaches the shape of a truncated icosahedron. The topology of cell arrangement remains but the curvature is increased at cells with five neighbors. Such configurations have also been found in \cite{Aland_MMS_2012}, which combines a surface phase-field crystal model with bending properties of the surface and a similar multiscale coupling to model the buckling instability in viral capsids. The obtained shapes deviate from the simple prolate or oblate shapes for $\kbend = const$ in Figure \ref{fig_Seifert}. Further reducing $V_r$ makes these truncated icosahedral shapes unfeasible and leads to further symmetry breaking and prolate-like shapes. The geometric axis of the prolate thereby emerges from one of the six symmetry axis determined by the positions of the cells with five neighbors. But also in these configurations cells with five neighbors are located in, or form, regions of high curvature, only the magnitude now differs and is largest at the cells at the symmetry axis of the prolate. This transition can be quantified by measuring the distance between opposed cells with five neighbors and the mean angle these axis form with each other, see Figure \ref{fig_quant_axis}. In these plots the deviation of one axis, the symmetry axis of the prolate, from all others is shown. The other five axis behave similar and only slightly decrease (length) or slightly increase (mean angle), as a geometric consequence of the volume and area constraints. For a sphere, $V_r = 1.0$, the values are equal and for the angle approximate the known quantity for a regular truncated icosahedron of $\arccos{\big(\frac{1}{\sqrt{5}}\big)} \approx 63.43^\circ$.

\subsection{Effect of activity}
\begin{figure}[!bht]
    \includegraphics[width=0.9\linewidth]{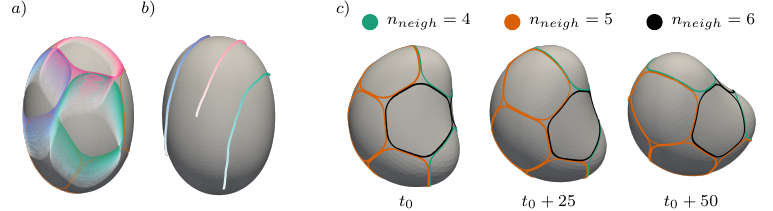}
    \caption{Collective rotation for $N=12$ and $v_0=0.16$. For $V_r=0.97$ (Outlines from subsequent timesteps shown in a), Trajectories shown in b)) the cells rotated collectively, without changing the shape of the surface. For $V_r=0.95$ the collective rotation of the cells leads to an rotation of the surface as shown in the snapshots in $c)$. }
    \label{fig_Rotation}
\end{figure}

In the next step we consider these equilibrium configurations and add activity by setting $v_0 > 0$. For a sphere, $V_r =1$, this resamples the situation considered in \cite{Happel_EPL_2022}, where collective rotation was discovered as an intermediate phase between the solid and liquid phase of the tissue if the number of cells is small and the activity not too large. 
We consider the case $N=12$ and a similar parameter regime but now for reduced volumes $V_r < 1.0$. Figure \ref{fig_Rotation} shows some of the results. For $V_r \gtrsim 0.96$ we obtain the same behavior as in \cite{Happel_EPL_2022}. Cells collectively rotate. Figure \ref{fig_Rotation} a) shows overlaid cell contours of three cells for different time instances indicating their movement. The corresponding trajectories of the centers of mass of these cells are shown in Figure \ref{fig_Rotation} b). Both indicate collective rotation on a fixed shape. The direction of rotation depends on the initialization of $\mathbf{e}_i$, which are initialized from a random uniform distribution. 

For $V_r \lesssim 0.96$ the situation changes, see Figure \ref{fig_Rotation} c). Instead of collective rotation on a fixed shape the results indicate that the shape rotates and the cells are transported with the shape. Similar phenomena of rigid body rotations have been observed in models for fluid deformable surfaces \cite{Olshanskii_PF_2023,  Nestler_PAMM_2023,Porrmann_PF_2024}. Such models consider the tangential motion of the cells as a continuous surface fluid  \cite{torres2019modelling,reuther2020numerical,krause2023numerical}. To further confirm these different behaviors we quantify the shape of the cells and the surface for both cases. The first is considered using surface Minkowski tensors \cite{Happel_IFB_2025}. These tensors characterize rotational symmetries of shapes embedded in curved surfaces and are extensions of classical Minkowski tensors in integral and stochastic geometry \cite{SchroderTurk2011Minkowski}. We here consider the normalized eigenvalues of the irreducible surface Minkowski tensors $\mu_p$ as a shape measure for the cells, see \cite{Happel_IFB_2025} for details. The index $p$ denotes the symmetry under rotations by $\frac{2 \pi}{p}$ with $p$ being an integer. Briefly, $\mu_p \in [0,1]$ with $\mu_p = 1$ for a regular geodesic polygon with $p$ vertices, which, however, might not exist on general surfaces. Surface Minkowski tensors are translation and scaling invariant, which on the surface translates to invariance under any flow generated by a Killing vector field and invariance under constant conformal changes of the metric of the surface, respectively. Furthermore, as also shown in \cite{Happel_IFB_2025}, they are robust against small perturbations of the shape of the cells or the surface. These properties allow to compare the shape of the cells on surfaces, even if they experience different geometric properties. As all cells, at least on the prolate in Figure \ref{fig_Rotation} a) and b), have five neighbors, we expect an established five-fold symmetry also for the individual cells. For computational issues concerning $\mu_p$ see Appendix.

In order to quantify the shape of the surface we consider global shape classifications using spherical harmonic-based principal component analysis, a method  researched in depth in the fields of $3D$-particle morphology and surface reconstruction \cite{Wei_JRMGE_2022,Zhou_IJNAMG_2017} as well as in computer graphics \cite{Kazhdan_ESGP_2003}. For this method the surface is expressed as function $\rSurf$ on the unit sphere $S^2$ following \cite{Zhou_IJNAMG_2017}, whereby the origin is used as a reference point. This requires a restriction to star-like shapes (with respect to the origin). Then every point on the surface can be uniquely associated to two angles $(\argtheta,\argphi)$ so that $\surfParam(\argtheta, \argphi)$ is the point of the surface that we reach when we go from the origin in the direction {$\begin{pmatrix} \cos(\argphi)\sin(\argtheta), \sin(\argphi)\sin(\argtheta),\cos(\argtheta)\end{pmatrix}^T$}. Note that the star like shape ensures the uniqueness of the association.
With this we define:
\begin{align}
\rSurf(\argtheta,\argphi)=||\surfParam(\argtheta, \argphi)||_2
\end{align}
As the surface is now described by a scalar function on the unit sphere and as the spherical harmonics are an orthonormal basis on the sphere we can now write this as 
\begin{align}
 \rSurf(\argtheta,\argphi)=\sum\limits_{l=0}^{\infty} \sum\limits_{m=-l}^{l} \qml{m}{l} \sphericalHarmonics{m}{l}(\argtheta,\argphi)
\end{align}
where we can calculate $\qml{m}{l}$ with 
\begin{align}
     \qml{m}{l}=\int_{S^2}\rSurf(\argtheta,\argphi)\overline{\sphericalHarmonics{m}{l}(\argtheta,\argphi)},
\end{align}
where $\overline{ \{...\}}$ denotes the complex conjugate. To characterize and analyze the surface only the coefficients with $l\leq 12$ are regarded as this turned out to be sufficient to describe the general shape of a surface \cite{Zhou_EG_2015}, see also Figure \ref{fig_Vis_SphericalHarmonics} in Appendix for a convincing example. To get a rotation invariant measure we follow \cite{Kazhdan_ESGP_2003} and introduce
\begin{align}
    \ql{l}=\sum\limits_{m=-l}^{l}(\qml{m}{l})^2
\end{align}
and for a scale invariant measure we consider 
\begin{align}
\qlnorm{l}=\frac{\ql{l}}{\sum\limits_{l=0}^{l_{max}} \ql{l}} \approx \frac{\ql{l}}{\int_{S^2}\rSurf(\argtheta,\argphi)^2 }
\end{align}
following \cite{Horn_JOSAA_1988,Kazhdan_ESGP_2003}.

\begin{figure}[thb]
    \centering
    \includegraphics[width=\linewidth]{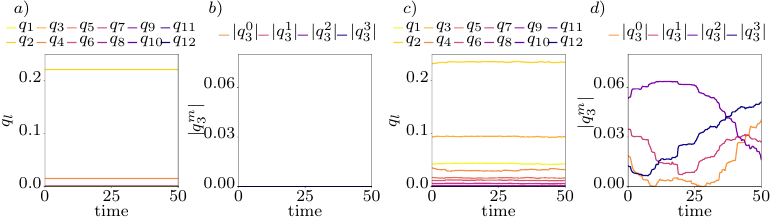}
    \caption{Surface shape quantifiers $q_l$ for $N=12$ and $v_0=0.16$. For $V_r=0.97$ $\ql{l}$ are shown in $a)$ and $|\qml{m}{3}|$ in $b)$, for $V_r=0.95$ $\ql{l}$ are shown in $c)$ and $|\qml{m}{3}|$ in $d)$. The axis scaling between $a)$ and $c)$ as well as $b)$ and $d)$ respectively is kept constant. }
    \label{fig_12_Cells_ShapeQuantifier}
\end{figure}

\begin{figure}[thb]
    \centering
    \includegraphics[width=0.8\linewidth]{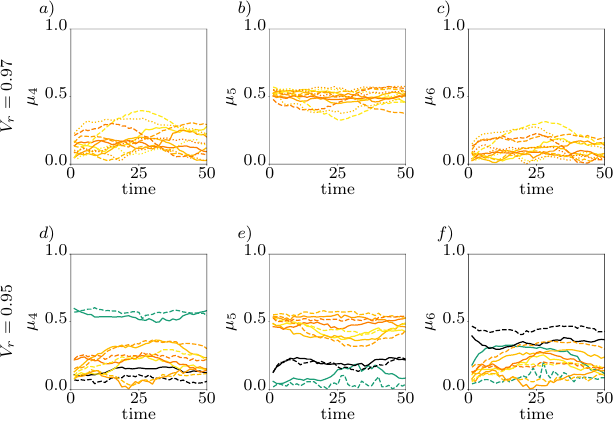}
    \caption{Cell shape measures $\mu_p$ for $N = 12$ and $v_0 = 0.16$. For $V_r=0.97$ $\mu_4$ are shown in a), $\mu_5$ in b) and $\mu_6$ in c). For $V_r=0.95$ the corresponding data is shown in d), e) and f), respectively. For each cell $\mu_p$ is shown over time. The line is colored according to the number of neighbors, whereby different linestyles are used to distinguish different cells with the same number of neighbors. Furthermore different shades of yellow and orange are used for cells with $5$ neighbors.}
    \label{fig_12_Cells_mu_p}
\end{figure}

Due to the rotational invariance of $q_l$ our shape measures stay constant, as can be seen in Figure \ref{fig_12_Cells_ShapeQuantifier} a) (for $V_r=0.97$, which corresponds to Figure \ref{fig_Rotation} a) and b))  and Figure \ref{fig_12_Cells_ShapeQuantifier} c) (for $V_r=0.95$, which corresponds to Figure \ref{fig_Rotation} c)). If we take a closer look at $\qml{m}{l}$, here for $l=3$, which are not rotation invariant, one can see that they are still constant for $V_r=0.97$ (see Figure \ref{fig_12_Cells_ShapeQuantifier} b)) while they are changing for $V_r=0.95$ (see Figure \ref{fig_12_Cells_ShapeQuantifier} d)). This confirms the postulated differences between Figure \ref{fig_Rotation} a) and b) (collectively rotating cells without changing the shape of the surface) and c) (collective rotation of the cells leading to rotation of the surface).

We now concentrate on the cell level and consider the irreducible surface Minkowski tensors $\mu_p$ to quantify the shape of the cells. Results are shown in Figure \ref{fig_12_Cells_mu_p}. For $V_r=0.97$ (see Figure \ref{fig_12_Cells_mu_p} a) - c)) and ${V_r=0.95}$ (see Figure \ref{fig_12_Cells_mu_p} d) - f)) we see a connection between the number of neighbors and the values of $\mu_p$. For example for $V_r=0.97$, where all cells have five neighbors, all cells have similar values for $\mu_5$, see Figure \ref{fig_12_Cells_mu_p} b), which are also significantly larger than the values for $\mu_4$ and $\mu_6$, see Figure \ref{fig_12_Cells_mu_p} a) and c), respectively. A similar behavior can be observed for $V_r=0.95$. Here the behavior depends on the considered cells. For cells with four or five neighbors (see Figure \ref{fig_12_Cells_mu_p} d) and e), respectively) the values for $\mu_4$ and $\mu_5$ are larger and stay roughly constant. Small shape changes, resulting from the random character of the incorporated activity, are expected. However, the overall behavior remains in a dynamic equilibrium not changing the global arrangement. 
At least for $V_r = 0.97$, where the cells collectively rotate on a prolate the cells experience different curvatures, see the selected cell trajectories in Figure \ref{fig_Rotation} b). While this modifies the shape of the cells it only has a minor influence on the five-fold rotational symmetry, see Figure \ref{fig_12_Cells_mu_p} b). Only for the lower values of $\mu_4$ and $\mu_6$ in Figure \ref{fig_12_Cells_mu_p} a) and c), respectively, fluctuations in time, which relate to the rotation, emerge.

\FloatBarrier
\subsection{Two-scale coupling for more cells}

\begin{figure}[!htb]
    \includegraphics[width=\linewidth]{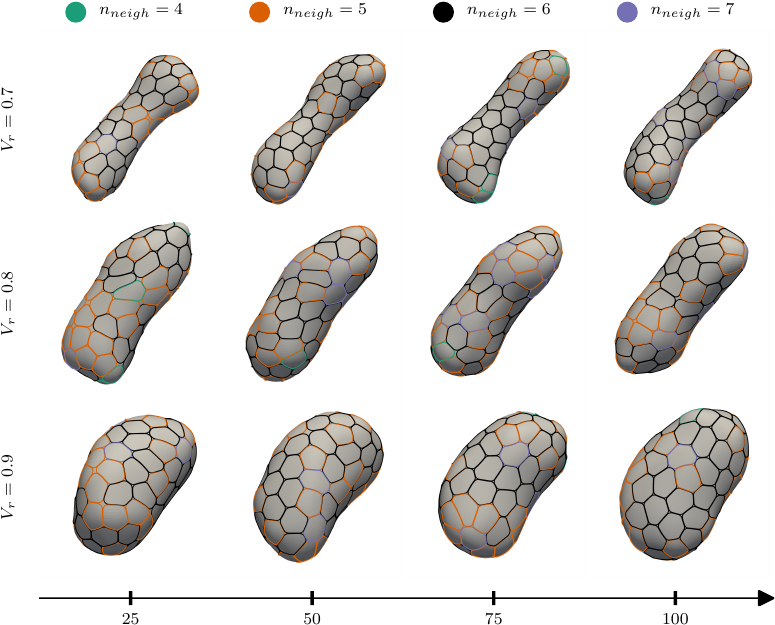}
    \caption{Shape evolution for $N=92$ and $v_0=0.2$ and different reduced volumes $V_r$. The cells are color coded according to the number of neighbors. The evolution is shown after an initialization phase. }
    \label{fig_Evol_Neigh}
\end{figure}

With the results of the considered problems with small numbers of cells, which resample known results and already demonstrate the possibility of new phenomena emerging from the two-scale coupling, we now turn to the full problem and consider $N = 92$. While still much lower than the envisioned problems in morphogenesis, the considered setting remains computational feasible and allows for statistical investigations. Figure \ref{fig_Evol_Neigh} shows snapshots of the evolution. The corresponding equilibrium shape for $\kbend = const$ and a random arrangement of cells was used as initial configuration. 

\begin{figure}[htb]
    \includegraphics[width=\linewidth]{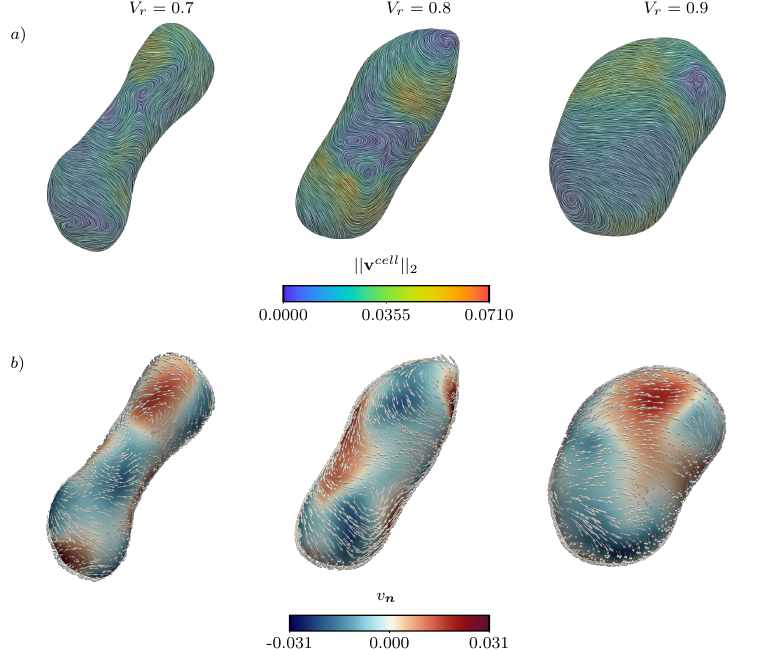}
    \caption{Global velocity field for simulations with $N=92$ and $v_0=0.2$ and different $V_r$ at time $t = 25$. a) LIC (line integral convolution) filter with color coding according to the magnitude of the global velocity field. b) Decomposed global velocity in its normal and tangential components. The magnitude of the normal part is color-coded. Thereby blue indicates movement inwards and red movement outwards. The direction of the tangential velocity field is shown by arrows, with length corresponding to the magnitude of the tangential velocity in log-scale.}
\label{fig_GlobalVelocityField}
\end{figure}

In all cases the shape evolves away from the equilibrium shape. Cells migrate and deform, due to interactions with other cells but also by experiencing different geometric properties of the evolving shape. Events can be identified where cells move relative to each other through structural rearrangements. This changes the number of neighbors and therefore the bending rigidity $\kbend$ which has an effect on the shape evolution, which is less regular than simulation results for continuous models, e.g., for active fluid deformable surfaces \cite{Porrmann_PF_2024}. For further comparison with such continuous approaches we also consider the surface velocity. To obtain a global velocity field we compute a velocity for each cell considering the centers of mass at the time instances $t$ and $t+5.0$. These velocities are summed up, whereby the rescaled phase-field of the respective cell is used as a weight. Afterwards the global field is smoothened with a linear interpolation kernel whereby the radius of this kernel corresponds to the radius of the cell. The corresponding velocity fields to Figure \ref{fig_Evol_Neigh} at time t = 25 are shown in Figure \ref{fig_GlobalVelocityField}. Two different ways to visualize the surface velocity are considered, the first focuses on the global velocity $\mathbf{v}^{cell}$ (Figure \ref{fig_GlobalVelocityField} a)) and the second decomposes it into its tangential and normal components $\mathbf{v}^{cell} = \mathbf{v}^{cell}_{tangential} + v_{\mathbf{n}} \mathbf{n}$ and visualizes both components (Figure \ref{fig_GlobalVelocityField} b)). As for simulations of fluid deformable surfaces \cite{reuther2020numerical,krause2023numerical} the normal and tangential velocity components are tightly coupled. Any shape change induces deformations of the cells and thus a tangential flow. Rearrangement of cells modifies the bending rigidity and induces local shape deformations thereby creating local gradients of mean curvature. Such gradients are of special interest as they are proposed as sources of active geometric forces \cite{AlIzzi_PRR_2023,Porrmann_PF_2024,Nitschke_PRSA_2025}. The generation of gradients in mean curvature thus might be an initialization for larger shape changes as observed in morphogenesis. 

\begin{figure}[!htb]
    \includegraphics[width=\linewidth]{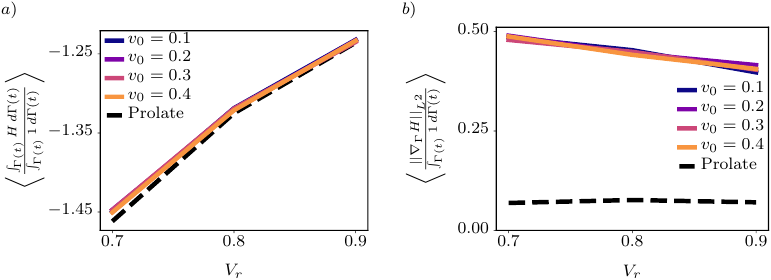}
    \caption{Local geometric properties. a) Averaged mean curvature $\left\langle \meanCurv \right\rangle$ and b) averaged mean curvature gradient $\left\langle \nabla_\Gamma \meanCurv \right\rangle$ as functions of the reduced volume $V_r$ for different activities $v_0$. The dashed line shows the values for the equilibrium prolate shapes from Figure \ref{fig_Seifert}.}
    \label{fig_DiagramCurvature}
\end{figure}

To analyze such evolutions and explore how the resulting shape changes depend on activity on the cellular scale we measure local and global geometric properties. We first compute the mean curvature $\meanCurv$ and the gradient of the mean curvature $\nabla_\Gamma \meanCurv$ at each point and time step and consider the average values. Figure \ref{fig_DiagramCurvature} shows the resulting values as a function of the reduced volume $V_r$ and for different values of the activity $v_0$ in comparison to the equilibrium prolate shape.   

As expected the averaged mean curvature decreases (negative sign for $\left\langle \meanCurv \right\rangle$) and the averaged gradient of the mean curvature increases for decreasing $V_r$, indicating the stronger deviation from a sphere. The strength of the activity has only a minor effect. While the averaged mean curvature essentially reproduces the properties of the equilibrium shape, the average mean curvature gradient strongly deviates. This results from the properties of the bending rigidity. Local variations in the bending rigidity lead to changes in mean curvature. 

\begin{figure}[!htb]
    \includegraphics[width=\linewidth]{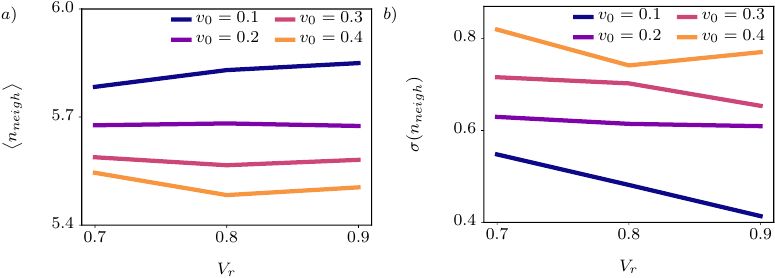}
    \caption{Number of neighbors: a) Mean number of neighbors $\left\langle n_{neigh} \right\rangle$ and b) standard deviation $\sigma(n_{neigh})$ as function of the reduced volume $V_r$ for different activities $v_0$.}
    \label{fig_Neighbor}
\end{figure}

\begin{figure}[!htb]
    \includegraphics[width=\linewidth]{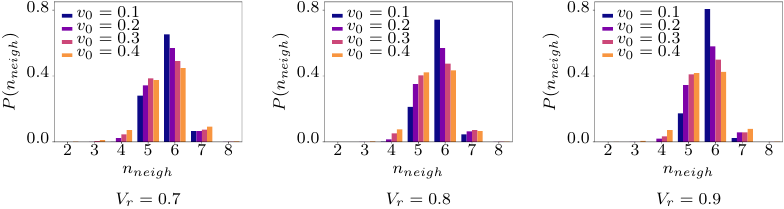}
    \caption{Distribution of the number of neighbors $n_{neigh}$ for different reduced volumes $V_r$ and different activities $v_0$.}
    \label{fig_Histogram_Neighbor}
\end{figure}

\begin{figure}[!htb]
    \includegraphics[width=\linewidth]{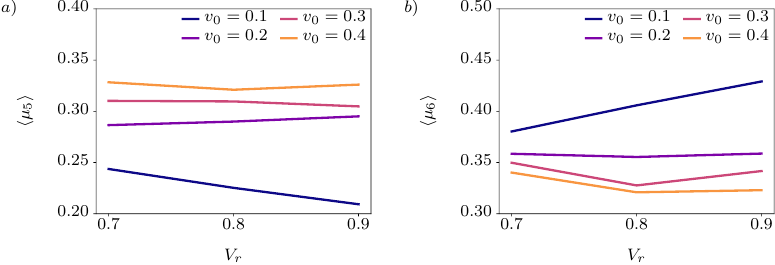}
    \caption{Averaged irreducible surface Minkowski tensors $\mu_p$. a) $p=5$ and b) $p=6$ for all cells as functions of the reduced volume $V_r$ for different activities $v_0$.}
    \label{fig_CellShape}
\end{figure}

Besides these averaged local geometric quantities we are also interested in a potential correlation between the mean curvature $\meanCurv$ and the number of neighbors of the cells $n_{neigh}$. We therefore compute the mean number of neighbors and its standard deviation. Both quantities are shown in Figure \ref{fig_Neighbor} as a function of the reduced volume $V_r$ and for different values of the activity $v_0$. The corresponding distributions are shown in Figure \ref{fig_Histogram_Neighbor} as histograms. While the values only slightly change with $V_r$ the mean number of neighbors decreases towards five with increasing $v_0$, which is associated with an increase in the standard deviation. In flat space the associated neighbor distributions are very robust. They are always centered at six and a broadening with activity is well known \cite{Wenzel_JCP_2019}, which is also observed here. A shift of the mean value towards five furthermore is consistent with results on a sphere \cite{Happel_EPL_2022}. The shift towards five and the broadening of the distribution leads to a strong increase in the number of topological defects and with it an increase in heterogeneity of the bending rigidity $\kbend$, which on average becomes weaker. The increase in heterogeneity with respect to neighbor relations correlates with a shift from dominating six-fold to five-fold rotational symmetry. This can be quantified by computing the irreducible surface Minkowski tensors $\mu_5$ and $\mu_6$, see \cite{Happel_IFB_2025}, for all cells and computing their average, which are shown in Figure \ref{fig_CellShape}. 

\begin{figure}[!htb]
    \includegraphics[width=\linewidth]{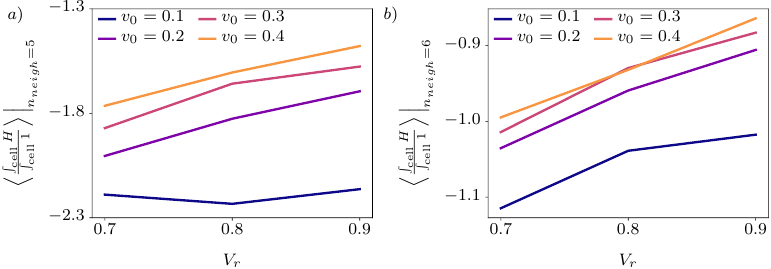}
    \caption{Local geometric properties at specific cells: a) Average mean curvature $\left\langle \meanCurv \right\rangle$ for all cells with five neighbors and b) average mean curvature $\left\langle \meanCurv \right\rangle$ for all cells with six neighbors as function of the reduced volume $V_r$ for different activities $v_0$.}
    \label{fig_NeighborCurvature}
\end{figure}

We next relate the number of neighbors $n_{neigh}$ and the mean curvature $\meanCurv$ to each other. Figure \ref{fig_NeighborCurvature} shows the average value of the mean curvature at cells with five and cells with six neighbors in a) and b), respectively. Other numbers $n_{neigh}$, while rarely present, do not allow for meaningful statistics and are therefore not discussed. Again the data is shown as a function of the reduced volume $V_r$ and for different  activities $v_0$. Both quantities consistently decrease (negative sign for $\left\langle \meanCurv \right\rangle$) with decreasing $V_r$ and increase with increasing $v_0$. However, the values for cells with five neighbors are significantly lower (negative sign for $\left\langle \meanCurv \right\rangle$) than the once for cells with six neighbors. The absolute difference between these values decreases with increasing $v_0$. 

At last we aim to quantify global shape changes. For this purpose we consider the shape measures $\qlnorm{l}$. We consider two quantities, the first considers the average change of the shape in a fixed time unit, here $0.125$, and the second the average change with respect to the initial prolate shape. These quantities read
\begin{align}
   \mbox{relative shape change} &= \frac{1}{n_{timestep}} \sum_{t^n} \sqrt{\sum_0^{l_{max}} (\qlnorm{l}(t^n+ 0.125) - \qlnorm{l}(t^n))^2} \\
   \mbox{absolute shape change} &= \frac{1}{n_{timestep}} \sum_{t^n} \sqrt{\sum_0^{l_{max}} (\qlnorm{l}(t^n) - \qlnorm{l}(0))^2}, 
\end{align}
where $n_{timestep}$ is the number of time steps and $t^n$ the n-th time instant. These quantities increase with decreasing reduced volume $V_r$, which can be explained by the enlarged shape space which is explored by the evolution. The dependency on activity $v_0$ is more subtle: The relative shape change increases with activity while the absolute shape change decreases. This corresponds to larger shape fluctuation on short time scales but less deviations from the equilibrium configurations over long time scales. Activity thus allows to explore the shape space but also to approach shapes close to equilibrium. A property which had also been found numerically for active fluid deformable surfaces \cite{Porrmann_PF_2024}. However, here it is also a consequence of the considered activity, which contains a random component. Furthermore the constraints on area and enclosed volume of the global shape restrict strong deviations. The last aspect considers the time evolution. While stronger activities lead to more heterogeneity and stronger tangential motion, the relaxation in normal direction is not directly effected. This might result in situations where the shape has not enough time to adapt to local changes in bending rigidity before neighbor relations are changed again. 

\begin{figure}[!htb]
    \includegraphics[width=\linewidth]{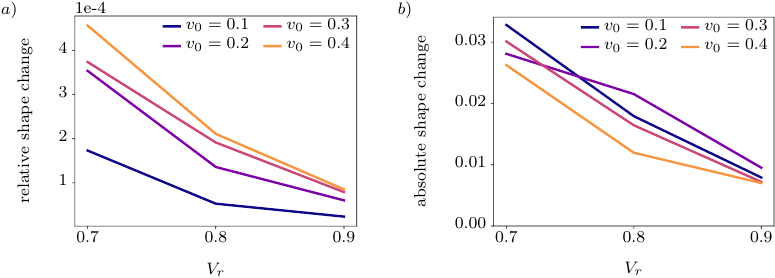}
    \caption{a) Relative shape change with respect to consecutive time instances and b) absolute shape change with respect to initial prolate shape for 
    $N = 92$ and the considered time period $t^n \in [10,99.775]$ as function of the reduced volume $V_r$ for different values of the activity $v_0$.}
    \label{fig_ShapeQuantifier}
\end{figure}

\section{Discussion}
\label{sec:4}

Large-scale motion in developmental tissue deformations emerges from collective tissue-mechanical properties of multicellular structures. Unfortunately these properties on the tissue scale are largely unknown, which limits predictive quantitative simulations of morphogenetic processes. As one step to overcome this limitation we have proposed a two-scale modeling approach in which each individual cell is resolved and the properties on the tissue scale do not need to be imposed but emerge from the interactions of the cells. This shifts the problem to providing mechanical properties of the cells. To construct a computationally feasible model requires various modeling assumptions. We have made use of scale separation to decouple processes on the cellular and tissues scale as much as possible and only consider the simplest possible model for a deformable cell. One of the coupling mechanisms is in the definition of the bending rigidity of the tissue, which locally depends on the topology of the cellular network. Being inspired by thin crystalline sheets, which buckle at defects, the bending rigidity is reduced if cells have less or more than six neighbors. Using advanced software tools, which essentially scale with the number of cells on parallel architectures, the features of the proposed two-scale model have been explored. 

For small numbers of cells and spherical shapes ($V_r = 1.0$), which resample epithelial acini \cite{Tanneretal_PNAS_2012,Wangetal_PNAS_2013}, we observed collective rotation, as in \cite{Happel_EPL_2022}. However, this changes if the reduced volume $V_r$ is reduced. Below some threshold the motion turns from a collective motion on a more or less stationary shape to a rigid body rotation, where the cells more or less remain stationary but collectively rotate the global shape. Activity on the cellular level thus not only leads to collective motion in form of tangential flow but also to shape deformations of the tissue. The resulting shapes have broken symmetries, which strongly depend on the topology of the cellular arrangement. These configurations show larger absolute values of the mean curvature at topological defects. For larger numbers of cells we have analyzed averaged quantities. 
We found global flow patterns, which show a tight coupling between the tangential $\mathbf{v}_{tangential}^{cell}$ and the normal velocity $v_\mathbf{n}$ components, reminiscent to fluid deformable surfaces \cite{torres2019modelling,reuther2020numerical}. An other finding was a strong increase in gradients of the mean curvature $\nabla_\Gamma H$ if compared with the equilibrium prolate shape. This results from the change in bending rigidity and can also be seen in classical two-component models on membranes with a phase-dependent bending rigidity, e.g., \cite{art:Baumgart2003,elliott2013computation,barrett2017finite,Bachini_JFM_2023}. However, here it appears on the cellular scale and might be a mechanism to initiate active geometric forces \cite{AlIzzi_PRR_2023,Porrmann_PF_2024,Nitschke_PRSA_2025}, which require gradients in mean curvature. Other findings are concerned with the number of neighbor distribution, which slightly broadens with decreased reduced volume $V_r$ and increased activity $v_0$ and also shifts towards five for the mean value if $V_r$ is decreased or $v_0$ is increased. We thus obtained an increase in heterogeneity and associated with this a weakening of the bending properties of the tissue.  
These neighbor relations correlate with the rotational symmetry properties of the cells, which have been computed using irreducible surface Minkowski tensors \cite{Happel_IFB_2025}. 

While all these results remain qualitative, they demonstrate the influence of the cellular scale on large scale tissue deformations. The approach provides a way to incorporate activity on the level it is generated. In principle this opens up possibilities to also consider sub-cellular processes, even genetic and mechano-chemical processes. However, modeling assumptions on the cellular scale are expected to have an  impact on the shape evolution. We here considered the simplest possible model for a cell. As already seen in flat space assumptions on the motility mechanism lead to different mechanical properties of the tissue \cite{Wenzel_PRE_2021}. Also extrinsic curvature effects, which lead to an alignment of the cells with principal curvature directions of the underlying surface \cite{PhysRevLett.129.118001,Happel_PRL_2024} have been neglected. Other strong assumptions are the constant cell size and to neglect cell proliferation. Quantitative descriptions of morphogenetic processes will certainly require model refinements in these directions. While this seems feasible for future investigations, the largest obstacle to overcome to enable simulations of interesting morphogenetic evolutions, such as the gastrulation process considered in \cite{munster2019attachment}, are limitations of the numerics. The computational effort to resolve the geometric properties in the required accuracy to obtain stable numerical schemes is huge \cite{hardering2023tangential,Bachini_JNM_2023,sass2023accurate,elliott2025fully}. Improvements in numerical analysis are certainly needed to reach comparable problem sizes. On the other side, the proposed model is one step towards more quantitative descriptions and asks for dedicated experiments with a moderate number of cells and within a controlled environment to parameterize and validate the two-scale model.

\FloatBarrier
\appendix
\section{Appendix}
\subsection{Illustration for shape classification}
\begin{figure}[htb]
    \includegraphics[width=\linewidth]{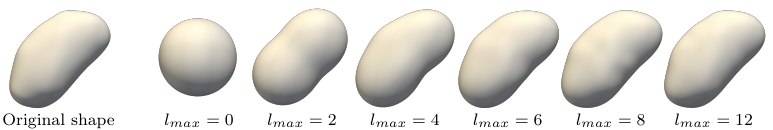}
    \caption{Visualization of the shape classification with spherical harmonics. On the left the original shape is shown, on the right we show reconstructions of the shape $\sum_{l=0}^{l_{max}} \sum_{m=-l}^{m=l} \qml{m}{l} \sphericalHarmonics{m}{l}(\argtheta,\argphi)$ for different values $l_{max}$. With higher $l_{max}$ the reconstructed shape resembles the original shape more and more.}
    \label{fig_Vis_SphericalHarmonics}
\end{figure}

Here we illustrate how the spherical harmonics can be used to reconstruct the original shape. Note that for a reconstruction the values $\qml{m}{l}$ need to be used, as we lose information going from $\qml{m}{l}$ to $\ql{l}$. Furthermore we use here values of $\qml{m}{l}$ which are not normalized to obtain the same scale as in the original picture. As can be seen in Figure \ref{fig_Vis_SphericalHarmonics} the approximation already reproduces the original shape qualitatively with $l_{max}=4$. Higher $l_{max}$ improve only small details of the overall shape and for $l_{max}=12$ differences are no longer visible, which suggests $l_{max}=12$ to be sufficient for our purpose.

\subsection{Simulation details}
In all simulations we use an adaptive grid whereby we always have at least seven grid points in the diffused interface of each cell. For simulations with $12$ and $32$ cells $50$ time units are regarded. For the simulations with $92$ cells $100$ time units are regarded. There the first $10$ time units are excluded from all evaluations to ensure that the system had enough time to deviate from the initial state. The parameters used in all simulations can be found in Table \ref{tab:Par_const}. For the simulations with $92$ cells we consider $v_0 \in [0.1,0.2,0.3,0.4]$.

Due to the diffuse interfaces of the phase-field approach a packing fraction of $100\%$ is not feasible. Therefore in all simulations a packing fraction of $94\%$ is used. This also resembles biological tissues as the importance of unoccupied spaces between cells was shown for example in \cite{kim2021embryonic}.

\begin{table}[htb]
    \centering
    \begin{tabular}{c|c|c}
    Parameter & Explanation &  Value \\
    \hline
    $\eps$ & width of the diffuse interface in the phase-field description &$0.01$ \\
    $\FacCa$& capillary number & $10.0$ \\
    $\FacIn$ & interaction number & $0.05$ \\
    ${a}_{rep}$& strength of repulsive interaction & $0.0625$ \\
    $a_{att}$ & strength of attractive interaction & $0.48$ \\
    $\tau_n$ & time step & $0.005$ \\
     $D_r$ & rotational diffusion parameter & $0.005$ \\
     $\kfac$& range of the cell dependent bending stiffness &$0.022$ \\
     $\klower$& minimum of the cell dependent bending stiffness & $0.008$
    \end{tabular}
    \caption{Parameters used in all simulations}
    \label{tab:Par_const}
\end{table}
\subsection{Calculation of irreducible surface Minkowski tensors}
For the calculation of $\mu_p$ we use the zero-levelset of the cells as the cell contour. The zero-levelset was extracted with the help of \textit{dune-tpmc} \cite{Engwer_ACM_2018}. As this algorithm is based on a marching cubes algorithm the contours sometimes contain grid cell shaped artifacts. To get rid of these the cell contours where smoothened by replacing every coordinate with the mean of seven coordinates. The calculation of $\mu_p$ is done as described in \cite{Happel_IFB_2025} and the implementation follows closely the one in \cite{SurfaceMinkowski-jl}.
\section*{Acknowledgments}
We acknowledge support by Maik Porrmann on the realization of the classical Canham/Helfrich problem and fruitful discussion with Simon Praetorius and Florian Roß on the implementation in AMDiS. This work was supported by the German Research Foundation (DFG) through the research unit FOR3013, ``Vector- and Tensor-Valued Surface PDEs,'' within project TP01, ``Numerical methods for surface fluids'' (project numbers VO 899/28-2) and TP05, ``Ordering and defects on deformable surfaces'' (project number VO 899/30-1). We further acknowledge computing resources at JSC under grant "MORPH" and at ZIH under grant WIR.

\bibliography{library_curvature}
\end{document}